# PlaceRaider: Virtual Theft in Physical Spaces with Smartphones

Robert Templeman[†,‡], Zahid Rahman[†], David Crandall[†], Apu Kapadia[†]

[†]School of Informatics and Computing
Indiana University
Bloomington, IN, USA

[‡]Naval Surface Warfare Center
Crane Division
Crane, IN, USA

September 27, 2012

**Abstract**

As smartphones become more pervasive, they are increasingly targeted by malware. At the same time, each new generation of smartphone features increasingly powerful onboard sensor suites. A new strain of 'sensor malware' has been developing that leverages these sensors to steal information from the physical environment — e.g., researchers have recently demonstrated how malware can 'listen' for spoken credit card numbers through the microphone, or 'feel' keystroke vibrations using the accelerometer. Yet the possibilities of what malware can 'see' through a camera have been understudied.

This paper introduces a novel 'visual malware' called PlaceRaider, which allows remote attackers to engage in remote reconnaissance and what we call "virtual theft." Through completely opportunistic use of the phone's camera and other sensors, PlaceRaider constructs rich, three dimensional models of indoor environments. Remote burglars can thus 'download' the physical space, study the environment carefully, and steal virtual objects from the environment (such as financial documents, information on computer monitors, and personally identifiable information). Through two human subject studies we demonstrate the effectiveness of using mobile devices as powerful surveillance and virtual theft platforms, and we suggest several possible defenses against visual malware.

# 1  Introduction

The computational power and popularity of modern smartphones (now owned by nearly one in every two American adults [31]) have created opportunities for sophisticated new types of mobile malware. One particularly dangerous strain is 'sensory malware' [30], which opens up a new and physically intrusive attack space by abusing the on-board sensors of a smartphone. These sensors give malware the ability to observe the physical enviornment around the phone, allowing attackers to go beyond simply stealing the electronic data stored on the phone to being able to intrude on a user's private physical space. For example, researchers have demonstrated novel attacks in which a smartphone's microphone can be used to 'hear' sensitive spoken information such as credit card numbers [30], while the accelerometer can be used to 'feel' vibrations to infer keystrokes typed on a computer keyboard near the phone [21].

In this paper we study the privacy implications of 'visual malware', which uses the smartphone camera to observe its unsuspecting owner's physical space. While several interesting visual attacks have been proposed recently, they require the attacker (or their specialized hardware) to be within visual range of the victim [17, 1, 2, 29, 23, 18]. Xu et al. propose visual malware for smartphones that captures and uploads video to the remote attacker [35] but leaves the processing of this data to the attacker.

We introduce a proof-of-concept Trojan called "PlaceRaider" to demonstrate the invasive potential of visual malware beyond simple photo or video uploads. In particular we show how PlaceRaider allows remote hackers to reconstruct rich three-dimensional (3D) models of the smartphone owner's personal indoor spaces through *completely opportunistic use* of the camera. Once the visual data has been transferred and reconstructed into a 3D model, the remote attacker can surveil the target's private home or work space, and engage in virtual theft by exploring, viewing, and stealing the contents of visible objects including sensitive documents, personal photographs, and computer monitors. Figure 1 illustrates how PlaceRaider can use images captured surreptitiously during normal use of the phone to generate 3D models of the user's environment in remarkable detail. Our tools allow the attacker to explore the model through 3D navigation, and then zoom into particular regions to examine individual images. PlaceRaider thus turns an individual's mobile device against him- or herself, creating an advanced surveillance platform capable of reconstructing the user's physical environment for exploration and exploitation.

***Research challenges.*** The cameras and other sensors on a smartphone can collect vast amounts of data very rapidly, presenting a deluge of data to the attacker. For example, a 10-megapixel camera can easily record 100 megabytes of data per minute if taken at high rates, while other sensors like accelerometers have sampling rates nearing 100 Hz. This means that after implanting a Trojan on a phone, an attacker has access to a rich stream of sensor data, but with this richness comes two major challenges. First, the large amount of data may overwhelm the storage or radio communication capacities of the mobile device. A challenge lies in reducing the quantity of data such that the amount of useless or redundant information in the dataset is minimized, while as much of the valuable information as possible is retained. We explore the research question of how additional sensor data (from the accelerometer or gyroscope) can be used to reduce the amount of visual information needed to produce accurate reconstructions.

Second, even after the visual data have been intercepted and transferred from the device, the attacker faces the problem of sifting through hours or days of images in order to glean information about the environment. We propose to employ sophisticated algorithms from the computer vision community to 'convert' large unstructured collections of noisy images into coherent 3D models that are easy to understand and navigate [8, 33]. However, the reconstruction techniques that build these models typically have been applied only to deliberately taken, high-quality imagery from cooperative users, with thoughtful composition, accurate focus, and proper exposure. In this paper we study whether reasonable 3D models can be reconstructed from opportunistically obtained images, and whether these models provide sufficient fidelity for an attacker to extract private information from the target's physical space.

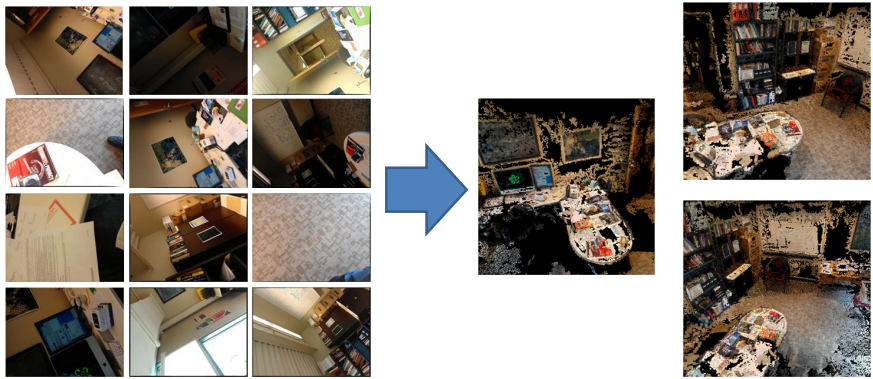

Figure 1: Illustration of our virtual theft attack. Photos (left) are taken surreptitiously by PlaceRaider using a smartphone's camera as its unsuspecting owner goes throughout his or her day. These unstructured opportunistic images are used by an attacker to generate a rich 3D model of the environment (right), providing contextual information about the environment and valuable spatial relationships between objects. These sample images and reconstructions are taken from the human subject studies we used to evaluate PlaceRaider, and provide an accurate illustration of our attack.

*Our Contributions.* We make the following specific contributions:

1. **Introducing an invasive strain of visual malware.** Previous examples of sensory malware have had narrowly targeted objectives (e.g., recovery of keystrokes or bank account numbers). PlaceRaider is the first example of sensory malware that threatens privacy in a more general manner using a combination of sensors, showing that through virtual theft a malicious actor can explore personal spaces and exploit or steal a plethora of sensitive information.

2. **Reconstructing spaces from opportunistic images.** We show that accurate 3D models can be created from opportunistically created photos taken by a smartphone camera as the device undergoes normal use. To our knowledge, ours is the first work to apply reconstruction algorithms to data of this type, since most work has studied thoughtfully taken photos (e.g. obtained from consumer photo-sharing sites like Flickr). To facilitate economical computation and reduce storage and network load, we develop heuristics using image quality metrics and sensor data that quickly identify the small fraction of images likely to contain valuable evidence and discard the rest.

3. **Developing tools to aid virtual burglary.** We develop and demonstrate a tool that allows an attacker to visualize and navigate a victim's space in 3D, allowing them to quickly hone in on areas that likely contain sensitive or private information and then retrieve targeted high-resolution images.

4. **Implementing and evaluating virtual theft.** We implemented PlaceRaider for the Android platform and evaluated our technique through two human subject studies. In the first study, images and sensor data are obtained in an indoor office from human participants during typical use of the device. Our study shows that high-quality reconstructions of indoor spaces using opportunistic data yields models with sufficient granularity to aid efficient exploration by malicious actors. Our second study evaluates how effectively attackers can understand the global structure of the dataset and their ability to 'steal' items from the 3D model.

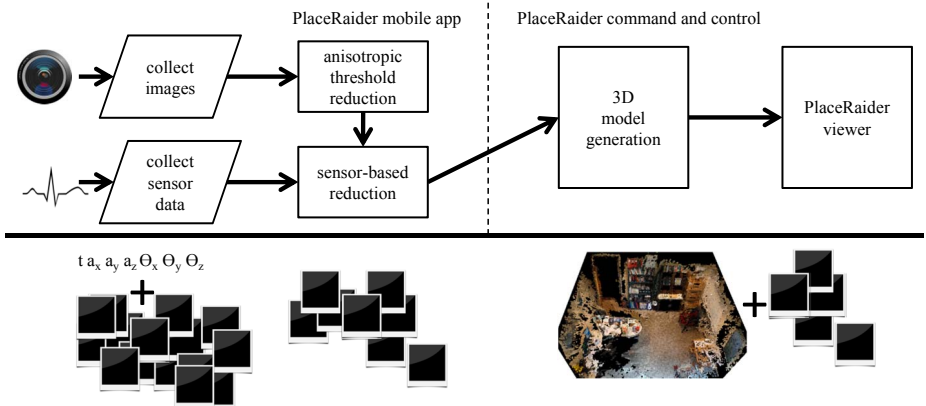

Figure 2: PlaceRaider architecture is depicted above. The data is collected using remote services on the mobile device. On-board preprocessing performs gross data reduction and packaging for transmission. The model is generated off-board and is navigated with a tool that allows exploration in a realistic 3D environment that provides an easy method to mine the image set. The malicious actor can easily extract sensitive information.

The remainder of the paper describes these contributions in detail. In Section 2 we provide a high-level description of our architecture, the adversary model and capabilities, constraints, and a concept of operation. Section 2.1 details the approach to collect and reduce data on the mobile platform, while Section 2.2 describes the model generation and data exploration methods that are necessary to exploit the dataset. Section 3 evaluates our data reduction techniques and the feasibility of virtual theft through human subject studies. We discuss the implications of our results along with limitations of such attacks, potential defenses and future work in Section 4, before discussing related work in Section 5 and concluding in Section 6.

## 2 Our Approach

The high-level PlaceRaider architecture is shown in Figure 2. A mobile device is infected with the PlaceRaider App, which we assume is embedded within a Trojan Horse application (such as one of the many enhanced camera applications already available on mobile app market places). We implemented PlaceRaider for the Android platform, creating 'remote services' that collect sensor data including images and acceleration and orientation readings. These remote services can run in the background, independent of applications and with no user interface. (We implemented on Android for practical reasons, but we expect such malware to generalize to other platforms such as iOS and Windows Phone.) The raw data is reduced and formatted before being transmitted to the PlaceRaider command and control platform. The 3D models are generated through this platform, where the burglar can explore and exploit the model and associated images.

To further illustrate how PlaceRaider works, we consider a detailed attack scenario. Alice often works from home, where she uses her mobile phone from time to time throughout the day for voice calls, responding to email, and Internet browsing. Alice's office at home has personal documents scattered on her desktop, including financial statements, phone numbers and a personal check. Her office walls have photos of her family and a wall calendar that shows her plans for the month (including a week-long, out-of-town vacation).

Alice does not know that her Android phone is running a service, PlaceRaider, that records photos surreptitiously, along with orientation and acceleration sensor data. After on-board analysis, her phone parses the collected images and extracts those that seem to contain valuable information about her environment. At opportune moments, her phone discretely transmits a package of images to a remote PlaceRaider command and control server.

Upon receiving Alice's images, the PlaceRaider command and control server runs a computer vision algorithm to generate a rich 3D model. This model allows Mallory, the remote attacker, to immerse herself easily in Alice's environment. The fidelity of the model allows Mallory to see Alice's calendar, items on her desk surface and the layout of the room. Knowing that the desktop surface might yield valuable information, Mallory zooms into the images that generated the desktop and quickly finds a check that yields Alice's account and routing numbers along with her identity and home address. This provides immediate value. She also sees the wall calendar, noticing the dates that the family will be out of town, and ponders asking an associate who lives nearby to 'visit' the house while the family is away and 'borrow' the iMac that Mallory sees in Alice's office.

***Assumptions.*** To make this type of attack possible, PlaceRaider makes a number of (weak) assumptions about the target and adversary:

- *Smartphone permissions.* The PlaceRaider App requires several access permissions from the Android operating system, in particular permission to access the camera, to write to external storage, and to connect to the network. Fortunately for PlaceRaider, all of these permissions would be needed for an innocent enhanced camera application, so asking the user for them is unlikely to arouse suspicion. PlaceRaider also needs permission to change audio settings (in order to capture photos without making an audible 'shutter' sound, as described in Section 2.1); this can also be easily justified to the user by, for example, advertising a feature that allows listening to music from an unrelated application without disturbing playback with 'annoying' shutter sounds. This coarse-grained permission is perceived as innocuous and even suggests it cannot cause harm to the user or phone [12] despite it controlling the mute state of the device. Even if the permissions could not be easily justified to the user, Felt *et al.* found that users often disregard permission warnings as most popular legitimate applications use various combinations of them [12]. No permissions are necessary to access high-fidelity sensor data from the accelerometers, gyroscopes, and magnetometers. Thus we expect few barriers to packaging PlaceRaider within an attractive camera app that will be downloaded by a large number of users.

- *Computation capability.* Generating the 3D models that we propose is very compute intensive, typically taking several hours of processing on a multi-core machine. Smartphone CPUs and GPUs are rapidly progressing (with modern ARM mobile processors approaching 70 gigaflops[1]) so that this computation may soon be possible on the phone itself, but for now we assume that generating the models is better done off-board, on the attacker's servers. To avoid having to transmit huge numbers of images to the servers, the PlaceRaider app applies lighter-weight image and sensor analysis to identify particularly promising images, removing redundant and uninformative images before transmitting the data to the PlaceRaider command and control center for model generation. Performing the 3D reconstruction on standard servers also allows us to use existing software tools that are not currently ported to the Android platform. Future work might include development of a mobile-centric application.

---

[1] http://www.anandtech.com/show/5077/arms-malit658-gpu-in-2013-up-to-10x-faster-than-mali400

- *Adversary model and objectives.* We assume that the adversary is interested in two major objectives: 1) *remote reconnaissance* of the the physical space through 3D reconstructions (e.g., to aid a physical burglary), and 2) *virtual theft* by allowing the attacker to peruse the victim's physical space for valuable information (e.g., sensitive personal and financial information). An attacker may be interested in only one of these objectives, but in either case we want PlaceRaider to provide an efficient method to quickly and accurately survey an area and extract information. Any adversary with modest computational resources who is able to create a suitable Trojan application and disseminate it through mobile app marketplaces can launch our proposed attack.

## 2.1 On-board data collection and filtering

We now describe the functionality of the PlaceRaider app, which collects images and sends a reduced set to the remote command and control server. The App has three main functions: to monitor orientation sensor data, to capture images, and to select a subset of images likely to contain valuable information for constructing a 3D model.

**Monitoring sensor data.** In the Android API *sensor* denotes functionality related to the accelerometers, gyroscopes, or magnetometers that a phone possesses. Surprisingly, as we mentioned in Section 2, no permissions are necessary to access such sensor data. We implemented a remote service that uses an event-driven listener that is called whenever sensor data changes. Android offers different degrees of refresh rates. We chose the fastest rate, which logs phone orientation and acceleration data at rates approaching 100Hz for the devices that we tested. Incoming data is tagged with a system time-stamp and logged.

**Surreptitiously taking images.** While Android cameras are capable of recording video or still images, we avoided video recording given the larger file sizes and additional power consumption that result. Using the Android camera without alerting the user or interrupting their interaction with the Android device poses a greater challenge than *sensor* data collection. The operating system imposes two requirements for use of the camera that try to prevent the camera from operating without alerting the user. The first is the photo preview feature, which provides a real-time display on the screen of what the camera is recording. However, this requirement can be circumvented relatively easily: the API requires that the preview be initialized with a Surface object prior to taking photos, but this can be satisfied by calling the method with a null object which in effect creates the surface with no visible area. The second requirement is that a shutter sound is played whenever a photo is taken. The playback of this sound cannot be prevented without root access to the phone, but again there is a simple workaround: we mute the speaker immediately before a photo is taken and restore the volume afterwards, so that playback of the shutter sound occurs but at a volume level of zero.

Modern Android devices provide a range of tunable camera parameters. While many of these phones boast camera resolutions of over 8 megapixels, such large images provide a large cost for computational handling and storage. We found that 3D models can be effectively reconstructed at much lower resolutions, as illustrated for one sample scene in Figure 3. Higher-resolution images do create more detail in the 3D model, but most of this detail is redundant information. We thus elected to record images at 1 megapixel, finding it to be a good trade-off between image size and model granularity. To obtain images that are as high quality as possible, we configure the camera for high shutter speeds in order to preserve as much image detail as possible even when the camera is moving. We configure the service to take photos at scheduled intervals via a timer, which we set to approximately one photo every 2 seconds in our implementation.

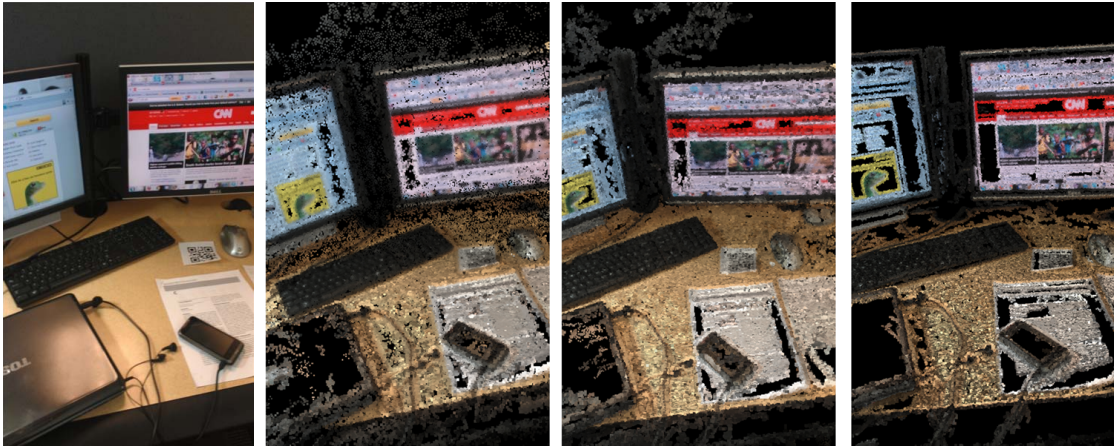

Figure 3: The leftmost image depicts an actual photograph taken at 6MP. The images to the right display perspectives of models generated from images down-sampled at .3MP, 1MP, and 4MP. It is clear that models generated from higher-resolution images offer greater granularity, but not so much to reveal more objects.

***Data reduction and transmission.*** Because images are taken surreptitiously and opportunistically, a large majority of the images are likely to be either of very low quality (e.g. completely dark photos while the phone is resting on a desk, or very blurry images when the phone is moving rapidly) or to be redundant (e.g. many duplicated pictures of exactly the same static scene). Such photos are not useful for the attacker, and thus we apply lightweight heuristics on the phone in order to identify and discard these uninformative images. The objective is to find a minimal subset of images that both preserves sufficient fidelity of the subsequently generated model and valuable information such that virtual theft can be realized. This data reduction conserves phone storage space and network bandwidth, and reduces the attacker's computational requirements for constructing the 3D models.

To filter out blurry and poorly exposed images that are unlikely to be useful to an attacker, we would like to compute a light-weight quality measure for a single image. Estimating quality metrics algorithmically is a difficult problem that has been studied extensively in the computer and human vision communities. Here we use an algorithm based on a measure of an image's anisotropy, which is the variance of the entropy of pixels as a function of the directionality [16]. We implemented this anisotropy-based quality method in PlaceRaider, allowing the malware to estimate the quality of captured images and discard images having a quality score less than some threshold value $T_q$.

After removing low-quality images based on the quality threshold we further reduce the image set by removing groups of successive images that are nearly identical. These groups of redundant images are typically quite common because in normal mobile device usage there are periods in which there is little physical movement of the phone. We use the phone's orientation sensors to detect times during which the phone is at rest or moving very little. The orientation sensor provides a time series of absolute 3D orientation vectors (specifying camera viewing direction parameterized by vectors on the unit sphere); let $\Theta_t = (\theta_x^t, \theta_y^t, \theta_z^t)$ denote the orientation vector at time $t$.

Taking finite differences, we can then approximate the angular velocity at time $t$ as

$$\Delta\Theta_t = ||\Theta_{t+1} - \Theta_t||.$$

Each image $i$ taken by our camera is associated with a time-stamp value $t_i$ that represents the image capture time logged by the system. We partition the images (ordered by timestamp) into groups of consecutive photos such that all images in the group have corresponding values of $\Delta\Theta_t$ less than a threshold $T_\theta$. Then we keep only the highest quality image (according to the anisotropy metric) from each of these partitions.

The architecture of Android devices introduces a practical complication to this approach, however: there is a lag between the hardware capture timestamp and the image timestamp due to the time required by the system to create the JPEG image. This means that sensor timestamps and image timestamps are misaligned by some unknown offset. For our test phones we estimated the lag to be nondeterministic but approximately normally distributed with a mean of 450 ms and standard deviation of 50 ms. To estimate the sensor data associated with an image, we take the mean of all the sensor samples which are within one standard deviation away from the mean of the actual sensor timestamp ($t_i' = t_i - 450$). This amount of lag will likely vary from phone to phone depending on hardware and operating system versions; in a real attack, the malware could calibrate itself by estimating phone motion using an optical flow algorithm [3] and compare this to sensor data, estimating the lag from a few minutes of data.

The effectiveness of this two-step reduction process depends on suitable values of $T_q$ and $T_\Theta$. In Section 3 we detail our method of finding suitable values for these parameters and an evaluation of the data reduction approach on empirical data.

## 2.2 Off-board reconstruction and exploitation

The final objective of PlaceRaider is to siphon images from a user's surroundings such that reconnaissance and virtual theft can occur in an efficient manner. While the stolen image set alone can provide valuable information, viewing images individually by hand is cumbersome and time-consuming with large image sets.

We propose the reconstruction of these images into a 3D model that gives an attacker a single coherent view of an environment, allowing her to navigate and explore the space in a natural way.

***Generating the model.*** 'Structure from motion' (SfM) is a well-studied problem in computer vision that generates a 3D model of a scene from a set of images taken from a moving camera. Recent work in SfM has introduced techniques that work on unstructured collections of images in which the motion parameters of the cameras are not known [8, 34]. The basic approach is to find points in images that are highly distinctive, so that 2D image points corresponding to the same 3D point in the original scene can be identified across different images. These (noisy) point correspondences induce constraints on the 3D structure of the scene points and the position and orientation of the camera that took each image. Performing 3D reconstruction thus involves solving a very large non-linear optimization problem to find camera poses and scene configurations that satisfy the constraints to the greatest degree possible. In this paper, we used an existing SfM software package, the Bundler toolkit,[2] to generate sparse 3D point clouds from our images [34]. SfM produces a relatively sparse reconstruction consisting of just the most distinctive scene points. To create a denser reconstruction, we follow up SfM with a multiview stereo algorithm, the Patch-based Multiview Stereo (PMVS) software of [15]. We made some changes to these tools to facilitate our work-flow and analysis, but the core algorithms and parameters were left unchanged.

---

[2] http://phototour.cs.washington.edu/bundler/

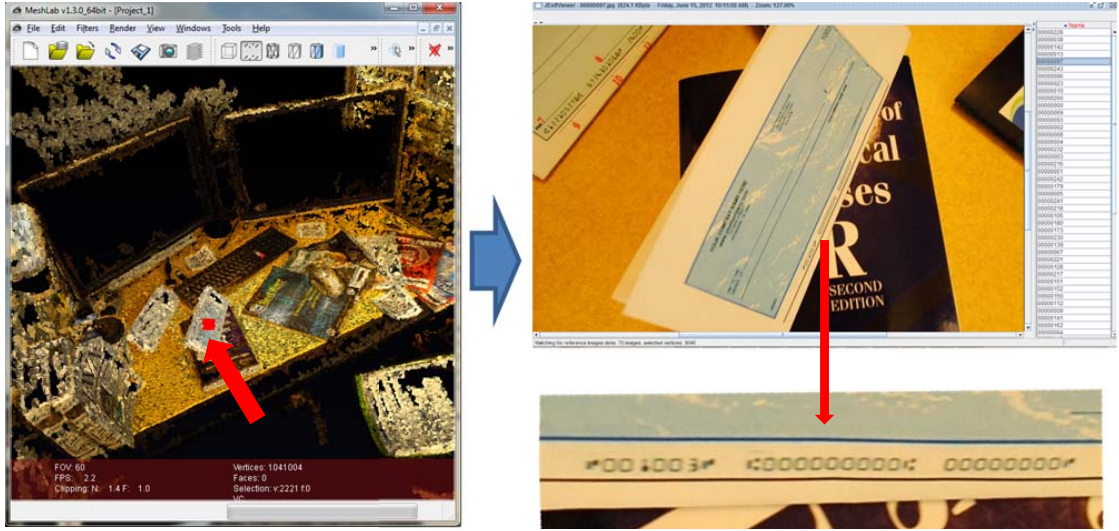

Figure 4: The PlaceRaider exploration tool. Users can navigate the 3D model using a point cloud viewer (left), and select some areas of particular interest (shown as red points). The PlaceRaider tool finds images that contribute to the selected region, presenting a list of matching images (top right); in this case, the image is detailed enough to clearly show the account numbers on a check (bottom right).

***Exploiting the model.*** The workflow just described generates a model (a 'point cloud') that is comprised of a collection of points, or vertices, each described with a three-dimensional location and color, and stored in the Stanford Polygon[3] (PLY) format. This format is understood by a variety of existing viewers. We used MeshLab[4] which is a popular open source viewer.

Exploration of even moderate-quality models generated with this approach yields reliable identification of gross features including walls, doors, windows, ceilings, pieces of furniture, and some smaller items. Higher-quality models that are built from commensurately high-quality images make it possible to read text on documents and identify small objects including keys and mobile devices. However, there are limitations, as even high-quality models will fall short of the detail contained in the images from which the models were based.

To explore the models more powerfully and facilitate the extraction of information, we developed a tool that allows users to select vertices in a 3D model and then shows the raw images from that region of interest in the model. To do this, we developed a MeshLab plugin that captures the selected vertices and then invokes a modified version of the open source image viewer JExifViewer.[5] Our modified viewer uses the PMVS patch file to extract the reference images associated with each of the selected vertices. These images are presented to the user in the order of coverage, showing the image that contains the greatest fraction of selected vertices first. This allows the attacker to easily identify images that have the greatest impact on the region of interest in the model. JExifViewer allows for convenient viewing of images with rotation and zooming capability to provide the user with a powerful interface for the presented images. Figure 4 demonstrates the appearance of a sample model from afar, zoomed in, and the extraction of high-resolution imagery.

---

[3]http://local.wasp.uwa.edu.au/~pbourke/dataformats/ply/
[4]http://meshlab.sourceforge.net/
[5]http://jexifviewer.sourceforge.net/

# 3  Evaluation

In this section we evaluate PlaceRaider and its potential to help carry out virtual theft. Our analysis explores three key questions. First, we test whether image sets captured opportunistically from smartphone cameras are sufficient to build 3D models of indoor spaces. To do this we conducted a human subjects study in which unsuspecting people carried out normal tasks on a smartphone that has been compromised with PlaceRaider. Second, we test whether the light-weight filtering techniques developed in Section 2.2, involving computing image quality scores and analyzing smartphone orientation sensor data, can be used to effectively reduce the size of the collected image set without sacrificing important visual information. Finally, we test the capability of an independent set of human participants to extract sensitive information from the models (about a physical environment they have never personally visited) using the PlaceRaider viewer utility.

## 3.1  Collecting opportunistic images

We designed a method to collect data comprised of images and sensor information from human participants using a smartphone with a Trojan application running in the background. The phone was an HTC Amaze running the unmodified Android 2.3.3 operating system. The experiment was conducted in the office of a university colleague who was not affiliated with our study; this room is of typical size and layout for an academic office (as shown in Figure 1). To preserve the anonymity of the owner of the test office, we removed objects having personally identifiable information, replacing them with other (simulated) sensitive features including personal checks, various barcodes, mock personal documents, and other personal effects. Lighting was typical overhead fluorescents with some sunlight through the window.

The PlaceRaider Trojan was installed on the phone and configured to take a photograph every two seconds. The camera was set with shutter priority so as to minimize motion blur and the resolution was set to 1 megapixel. The study consisted of 20 separate data collection events in the same office environment, using 20 participants recruited from a university campus. The test protocol led the participants through a series of activities designed to be representative of typical smartphone use, including Internet browsing, talking on the phone, and using basic applications. Participants were unaware of the nature of the study or the details of what data were being collected. (As described in Section 2.1, the PlaceRaider application disabled the Camera preview and shutter sound so that the user was not aware of our sensory collection effort.) In accordance with Institutional Review Board procedures, the participants were debriefed on the details of the study afterward.

As the study was self-paced, collection time varied by participant. Moreover, the PlaceRaider application occasionally stopped taking photos and sensor measurements and restarted using a watchdog timer, which led to further variation in the size of image sets (ranging from 115 to 1,520 images, with a mean of 864). For 4 of 20 test sets, sensor data was missing due to technical problems with the app. Nevertheless, all 20 datasets are useful to our study because missing data and degradation in observation quality is reflective of the real world. The first column of Table 1 summarizes the data collected by our study.

## 3.2  Generating 3D models

Once the data was collected from the human subjects study, we generated 3D models from each of the 20 raw image sets using the Bundler and PMVS software, as described in Section 2.2, and then evaluated the quality of the resulting 3D models. Unfortunately, quantitatively measuring the accuracy of 3D models is notoriously difficult unless ground truth is available in the form of high-fidelity laser scans [8]. We do not have such ground truth, so we instead evaluated the models subjectively. To do this, we identified several important features of the original scene, and then for each 3D model manually evaluated whether each feature had been faithfully constructed. Our features were: 4 walls, 1 countertop, 1 door, 3 chairs, 1 window, 1 bookshelf, 1 white board, 1 armoire, 2 posters, 2 LCD monitors, 2 desks, and a Dell computer box.

Table 1: Summary of image datasets collected by 20 human participants and the 3D models that were obtained, using both the raw image sets ($X_n$) and reduced sets ($X'_n$). The Pre-PMVS columns show the images input to the sparse 3D reconstruction algorithm, PMVS shows the number of images used by the dense multi-view stereo algorithm, while model quality score is the number of features visible in the model (where higher scores are better). TID's marked with '*' denote tests where reduction was performed without the sensor-based method.

| TID | ‖pre-PMVS image set‖ | | | ‖PMVS image set‖ | | | ‖model vertices‖ | | | model quality score | | |
|---|---|---|---|---|---|---|---|---|---|---|---|---|
| $n$ | $X_n$ | $X'_n$ | change | $X_n$ | $X'_n$ | change | $X_n$ | $X'_n$ | change | $X_n$ | $X'_n$ | change |
| 1 | 1365 | 244 | -82.1% | 187 | 46 | -75.4% | 142k | 17k | -88.1% | 2 | 0 | -2 |
| 2 | 1305 | 243 | -81.4% | 195 | 69 | -64.6% | 99k | 51k | -48.2% | 4 | 10 | 6 |
| 3 | 1214 | 432 | -64.4% | 773 | 362 | -53.2% | 813k | 607k | -25.3% | 19 | 19 | 0 |
| 4 | 778 | 126 | -83.8% | 21 | 14 | -33.3% | 33k | 26k | -21.5% | 3 | 4 | 1 |
| 5 | 449 | 39 | -91.3% | 207 | 4 | -98.1% | 192k | 0k | -100.0% | 3 | 0 | -3 |
| *6 | 162 | 83 | -48.8% | 82 | 70 | -14.6% | 60k | 48k | -19.3% | 3 | 1 | -2 |
| 7 | 974 | 490 | -49.7% | 370 | 198 | -46.5% | 393k | 271k | -31.1% | 10 | 8 | -2 |
| *8 | 446 | 254 | -43.0% | 63 | 61 | -3.2% | 57k | 71k | 24.2% | 4 | 6 | 2 |
| 9 | 719 | 124 | -82.8% | 289 | 61 | -78.9% | 206k | 142k | -30.9% | 11 | 15 | 4 |
| 10 | 541 | 92 | -83.0% | 87 | 21 | -75.9% | 56k | 32k | -42.9% | 5 | 5 | 0 |
| 11 | 1520 | 353 | -76.8% | 790 | 251 | -68.2% | 796k | 546k | -31.4% | 20 | 20 | 0 |
| 12 | 1378 | 157 | -88.6% | 99 | 5 | -94.9% | 83k | 2k | -97.2% | 4 | 0 | -4 |
| *13 | 1445 | 729 | -49.6% | 444 | 353 | -20.5% | 354k | 399k | 12.8% | 19 | 18 | -1 |
| 14 | 1059 | 150 | -85.8% | 2 | 2 | 0.0% | 0k | 0k | 0.0% | 0 | 0 | 0 |
| *15 | 846 | 558 | -34.0% | 162 | 160 | -1.2% | 86k | 99k | 15.4% | 4 | 6 | 2 |
| 16 | 446 | 62 | -86.1% | 58 | 7 | -87.9% | 11k | 21k | 86.3% | 0 | 3 | 3 |
| 17 | 414 | 42 | -89.9% | 47 | 11 | -76.6% | 7k | 0k | -100.0% | 0 | 0 | 0 |
| 18 | 762 | 162 | -78.7% | 36 | 49 | 36.1% | 21k | 68k | 217.1% | 0 | 1 | 1 |
| 19 | 115 | 17 | -85.2% | 7 | 2 | -71.4% | 1k | 0k | -100.0% | 0 | 0 | 0 |
| 20 | 1343 | 293 | -78.2% | 229 | 86 | -62.4% | 209k | 145k | -30.6% | 13 | 11 | -2 |
| $\mu$ | 864.05 | 232.50 | -73.2% | 207.40 | 91.60 | -49.5% | 181k | 127k | -20.5% | 6.20 | 6.35 | 0.15 |
| $\sigma$ | 447.28 | 193.22 | 17.9% | 230.43 | 113.68 | 37.5% | 240k | 184k | 73.9% | 6.74 | 6.93 | 2.43 |

We first ran Bundler and PMVS on each of the 20 raw image sets, without performing the image filtering proposed in Section 2.2. The $X_n$ columns of Table 1 present these results. We observe that the size of the reconstructed models varies across users, from over 800,000 3D points in the largest model to 0 points in the smallest model (which indicates that the 3D reconstruction failed to solve for any of the structure of the scene), with a mean of about 180,000 points. The subjective quality scores also varied significantly, from a score of 20 (indicating a perfect reconstruction according to our criteria) to 0 (again indicating that no useful model was generated), with a mean score of 6.2. Moreover, 30% of the models scored higher than 10, indicating that a majority of the scene was successfully reconstructed.

These results suggest that faithful 3D models of a space can often be generated from opportunistically captured images. This is a somewhat surprising result because most Structure from Motion approaches were designed for use with deliberately composed images. For example, some approaches assume that images have a low in-plane rotation twist angle (i.e. were taken such that the camera was level with respect to the ground) [8]. As Figure 1 shows the images used in our study are often of very low quality, with high twist angle, motion blur, poor focus, and random composition. The accuracy of the test models generated from these photographs demonstrates the ability of the feature detection and feature-key-matching algorithms to contribute to model generation despite the poor photography inherent to opportunistic image collection.

*Reducing the image sets.* As described in Section 2.1, we also investigated whether we can reduce the size of the photo collections by filtering out low-quality images on the smartphone itself in order to prevent having to store and transmit unnecessary images. This filtering uses light-weight computation that attempts to remove low-quality images using the anisotropic metric and redundant images by using the camera's orientation sensor to detect periods of low camera movement.

These techniques each require a threshold (the minimum quality score $T_q$ and the maximum change in orientation direction $T_\theta$), which intuitively trades off between the size of the image set and the quality of the resulting model. To estimate reasonable values for these thresholds, we took the model for one of the test subjects (#11, chosen because it had the largest number of images), and then examined the 730 (of 1,520) images that were not used by the 3D reconstruction algorithm (either because they were discarded as outliers or because they contained redundant information). We found that most images with quality scores $q$ less than 0.0012 were discarded by the PMVS process, so we set $T_q$ to this value. Applying this threshold to image set #11 resulted in a 50% reduction in the number of images compared to the raw image set and eliminated 80% of the images that would have been discarded by 3D reconstruction. For the threshold on orientation angle changes, we conservatively set $T_\theta = 8°$ based on the fact that the 3D reconstruction algorithm recommends at most 15° separation between camera viewing directions [32]. This process further reduced the image set for test #11 by 26.7%.

Having estimated these thresholds based only on one image set, we then applied the same two filtering techniques with these same thresholds on all 20 datasets, and then used Bundler and PMVS on these reduced datasets. The $X'_n$ columns of Table 1 show the results. On average the filtering removes about 73% of images from the raw dataset, reducing the mean photo collection size from 864.05 to 232.5 across the 20 test subjects. The number of points in the resulting 3D models also decreases, from about 181,000 to 127,000, but the model quality scores remain almost the same (and actually very slightly *increased* by an average of 0.15 features). Thus in spite of great reductions in the size of raw image datasets, the quality of the models were generally not compromised. This means that PlaceRaider can store and transmit only about 27% of the images collected and still provide an attacker with the same degree of information about the environment.

### 3.3 PlaceRaider model and viewer evaluation

The results in the last section suggest that 3D models produced by PlaceRaider could provide useful surveillance information about a space, but these evaluations were conducted by only one of the authors and are thus not necessarily unbiased. For a more objective test of how much an attacker unfamiliar with an environment could learn from our 3D models, we developed a visualization tool and conducted a second human subjects study in which we measured the ability of human participants to conduct virtual theft. As we described in Section 2.2, this visualization tool was created by adding a plug-in to MeshLab, an open-source tool for viewing 3D models. This tool allows the user to navigate the 3D point cloud model and to select particular points of interest. Meshlab then displays the raw images used to reconstruct that part of the scene.

Table 2: Summary statistics for the coarse feature identification task by human participants. The mean and standard deviation of counts are delineated by search method and feature type. The *coarse eval score* is the sum of listed identifiable features (13) subtracted by the absolute values of individual identification errors, including both the number of features that were not cataloged as well as cataloged features that were not present in the actual environment.

|  | walls | floor | ceiling | doors | windows | desks | tables | chairs | coarse eval score |
|---|---|---|---|---|---|---|---|---|---|
| number of features in space | 4 | 1 | 1 | 1 | 1 | 1 | 1 | 3 | 13 |
| raw image browsing $\mu_{n=8}$ | 3.75 | 2.50 | 1.50 | 3.75 | 1.25 | 2.13 | 1.13 | 5.00 | 0.75 |
| raw image browsing $\sigma_{n=8}$ | 2.82 | 1.31 | 0.93 | 1.67 | 1.04 | 1.64 | 0.99 | 2.62 | 5.95 |
| model navigation $\mu_{n=10}$ | 3.90 | 1.00 | 1.00 | 1.30 | 0.50 | 1.00 | 0.90 | 2.20 | 11.20 |
| model navigation $\sigma_{n=10}$ | 0.32 | 0.00 | 0.00 | 0.48 | 0.53 | 0.00 | 0.32 | 0.63 | 1.32 |

We presented the largest of our datasets (#11) to two groups of human participants (none of whom who were familiar with the physical office that had been modeled). The first set of participants ($N_1 = 8$) navigated the raw image set (1,520 images), without access to the model, by simply examining the images one-by-one. The second set of subjects ($N_2 = 10$) had access to a 3D model generated from the reduced set of images (353 in all) using our method described in Section 2.2. We asked each group of participants to perform two tasks, a coarse search for basic information about the space, and a fine search for specific sensitive details:

- *Coarse Search.* In the Coarse Search phase, the participant was told to catalog the presence of unique objects or features from a defined list, using either the set of individual images or the single model. These features are listed in Table 2. The objective of this phase was to assess the ability of participants to understand the context and global structure of the physical space. In this phase an objective score is assigned based on the ability of participants to catalog the features. The 'coarse eval score' represents how closely the participants were able to identify the 13 features in the room, and was calculated as the sum of listed identifiable features (13) subtracted by the absolute values of individual errors. These errors can be due either to under-counting (e.g. claiming one wall when four are in the room) or over-counting (e.g. claiming nine walls when only four are in the room).

- *Fine Search.* In the Fine Search phase, the participant was asked to act like a burglar looking for valuable information. The participant was given a list of items to look for, including financial documents, barcodes and QR codes, and other private information that have been planted in the original test space. One group of participants performs this search with solely the raw image set, while the other group of participants browses the model and queries regions of interest for targeted images using the utility that we developed and describe in Section 2.2. To motivate the participants to find as much information as possible, the participants were compensated by a cash payment that was commensurate with the number of objects that they successfully identified. An objective score for this phase was assigned to each participant based on the ability to extract sensitive information from the image list that they provided.

Each phase was preceded by a training activity in which the participant was presented with a training dataset to familiarize himself with our user interface and the process of information search.

**Results.** Table 2 presents the results for the Coarse Search phase. The group using the 3D model scored 11.2 on average with a standard deviation of 1.32 features. The group using only the raw image set scored much lower, at about 0.75 (with standard deviation of 5.9). The difference in means between the two groups was statistically significant ($p < 0.002$ according to Welch's t-test) and demonstrates that coarse feature identification and the aggregate relationship among an image set is greatly aided by use of a 3D model. An example of this is attempting to count the number of unique walls in our large collection of images; using the model it is obvious that the structure is a single room with four walls, but it is much more difficult if one must analyze over a thousand images individually (and subjects counted anywhere from 1 to 9 walls during this evaluation). Since many indoor walls are featureless surfaces, two different images of the same wall, under different lighting conditions, viewing angles, and positions, may be easily interpreted as images of two *different* walls without the context of a 3D model.

Table 3: Summary statistics for human participant performance on the fine feature identification task. The mean and standard deviation of counts are delineated by search method and feature type. The *fine eval score* is the sum of identifiable features (14) presented by the individual subjects.

|  | checks | bar codes | docu- ments | certif- icates | monitor | white board | desk surface | photo | fine eval score |
|---:|:---:|:---:|:---:|:---:|:---:|:---:|:---:|:---:|:---:|
| number of features in space | 2 | 3 | 3 | 2 | 1 | 1 | 1 | 1 | 14 |
| raw image browsing $\mu_{n=8}$ | 0.80 | 0.95 | 1.20 | 0.85 | 0.50 | 0.40 | 0.10 | 0.40 | 5.20 |
| raw image browsing $\sigma^2_{n=8}$ | 0.67 | 0.72 | 0.67 | 0.34 | 0.47 | 0.52 | 0.32 | 0.52 | 1.21 |
| model navigation $\mu_{n=10}$ | 1.00 | 0.25 | 1.00 | 1.00 | 0.63 | 0.63 | 0.00 | 0.44 | 4.94 |
| model navigation $\sigma^2_{n=10}$ | 0.60 | 0.46 | 0.60 | 0.65 | 0.44 | 0.52 | 0.00 | 0.42 | 2.15 |

Table 3 presents the results for the Fine Search phase. In this phase, the group using the model successfully identified 5.20 or 37.1% of the listed sensitive features on average, while the group using the raw image set found 4.94 or 35.3% of the sensitive features. This difference is not statistically significant, and thus we found no penalty or benefit in using the model to search for fine features using our PlaceRaider navigation utility. Viewed alternatively, using the model and the greatly reduced image set, there is no observable loss in the ability to extract fine features using our PlaceRaider implementation as compared to the raw data that is available from a mobile camera. Nevertheless the model is of clear benefit for spatial awareness, and we posit that it will perform much better for supporting navigation and virtual theft in larger spaces or multiple rooms.

## 4 Discussion

We first discuss how PlaceRaider can be improved from an attack perspective and then we discuss various possible defenses against such visual malware.

### 4.1 Improvements to the attack

***Targeted PlaceRaider.*** While providing a very general and powerful surveillance capability, our implementation of PlaceRaider relies solely on human vision for the extraction of detailed valuable information. A desired capability might include autonomous (machine) identification and extraction of sensitive data. Existing computer vision methods in object recognition and image matching [20] could be employed to look for a pre-defined set of objects of interest, as could optical character recognition or bar code recognition to glean text and other sensitive data from the scene. In doing this an interesting question is whether the 3D context provided by our reconstructions could improve object recognition over baseline algorithms that simply scan through the raw images.

***Improving PlaceRaider data reduction.*** While we were able to demonstrate successful data reduction averaging more than 73%, many of the images in the reduced sets still had poor quality and high redundancy. There are at least two possible potential methods to improve data reduction using orientation and acceleration data. The first involves utilizing the acceleration data that is already logged, but not utilized, by the PlaceRaider app. If proper signal processing were applied to the acceleration data, it might be possible to infer instantaneous movement at the time of image capture. Such a capability would allow velocity thresholds to be applied to images, ignoring images likely to have significant motion blur. A second method to improve data reduction would be removal of redundant images at the global level. Our current implementation removes redundant images by looking only at local temporal neighbors. An alternative approach would be to use the camera gyroscopes to record where the camera has pointed before, and to take photos only when it is oriented towards new, unseen directions. Such approaches will become more feasible as smartphones incorporate increasingly accurate sensors.

While the anisotropic quality metric was effective at reducing the size of our image sets, manual inspection of the subsets revealed regular occurrences of grossly under- or over-exposed images. An improvement of our reduction technique could target luminance distributions of images, thus targeting and discarding additional poor quality images. We leave exploration of this and the other suggested data reduction improvements for future work.

***Improving the 3D models.*** In the past, large-scale structure-from-motion algorithms typically have been used for outdoor scenes using images purposefully taken by photographers. In Section 3 we showed that Bundler and PMVS successfully generated models on our opportunistically collected images of indoor spaces. This new domain raises interesting computer vision questions about how to improve 3D model generation with more difficult image collections. One possible direction is to explore alternative algorithms for feature identification and comparison that would work better in indoor spaces (where illumination is more variable and distinct features are less common because of the many plain surfaces). It would also be interesting to use recent work that uses stronger assumptions about a scene, for example that the scene has many flat planes, to produce more accurate models for indoor scenes [14]. We leave this exploration for future work.

Another interesting direction is to understand why the datasets from some of our 20 test subjects produced much better models than others. In some cases this can explained simply by the size of the datasets, but in other cases there are clearly other unknown factors at work. These may relate to the lighting conditions in the room at different times of the day, or to the smartphone usage habits of different people (e.g. the amount they move while using the phone, or how often their fingers accidentally cover the smartphone lens).

## 4.2 Defenses

We have shown that virtual theft is a compelling threat to privacy. Visual malware is a nascent area that requires more research. While existing work has established visual attacks — we discuss some in Section 5 — PlaceRaider introduces a threat that requires a different defense. While evaluation of defenses is outside of the scope of this paper, we offer these possible solutions for seeding future work.

***Permissions and defense.*** We detailed the set of Android permissions needed by PlaceRaider in Section 2. Previous work has shown that permission sets can be partitioned by splitting malware functionality and permissions among multiple applications and using covert channels to communicate [30]. Such covert channels are limited in channel capacity and preclude their use for transmitting image or other sensor data between applications. This limitation allows us to safely assume that PlaceRaider can only be employed when the user accepts the application permissions at installation. Static permission checking can be performed automatically, but again, since common camera apps require similar permissions, we do not expect such mechanisms to provide a successful defense [24, 11].

This points to the obvious recommendation that users should be cautious about what they install, and download apps only from trusted software developers. In general, users should be suspicious of apps that have access to the camera and the Internet.

***Persistent shutter sound.*** We described in Section 2.1 the measures that Android devices take to 'require' a shutter sound when taking photographs and our trivial, but effective, method to circumvent this by muting the speaker. An inability to prevent the shutter sound would foil PlaceRaider's attempts to control the camera without detection. A persistent shutter sound, possibly with dedicated hardware that is impervious to overrides by software, is a recommended approach that would provide a complete defense against PlaceRaider. This obviously requires initiatives by hardware manufacturers.

***Permissions for sensors.*** Current Android and iOS operating systems require no permissions to access acceleration or gyroscope data despite the high rates of collection and accuracy that they possess. An explicit permission to use the sensor resources would not prevent virtual theft, as PlaceRaider only uses this data to further reduce the image set, but would increase the security of mobile devices; existing work has demonstrated attacks using accelerometers to analyze keystrokes, for example [21, 22, 27]. While we offer this as a defense, it would only minimally — if at all — preserve privacy, as there would be no prevention of the collection and transmission of images.

***Hardware protections.*** One possibility is to enhance the operating system architecture to allow photographs and video to be captured only when a physical button is pressed. Such a mechanism may prevent surreptitious capture of photographic data, but may prevent novel uses of the camera by legitimate applications.

***PlaceRaider detector.*** There is no logical motivation for users to intentionally take poor-quality photos that have any combination of improper focus, motion blur, improper exposure, or unusual orientations/twist. Just as PlaceRaider can use methods to parse images to maximize quality, the Android operating system can potentially assess each photo at time of capture, or over longer periods of time, and ask the user to confirm that they desire these images. Such an approach might frustrate PlaceRaider attempts to perform long-running image collection.

PlaceRaider differs from most legitimate camera applications in that individual photos are taken without input from the user interface. The API could possibly be modified such that certain security-related methods would require invocation linked to user input. This would allow a total defense against PlaceRaider, as attempts to take photos in a stealthy manner would be thwarted. We leave exploration of these approaches, including attempts to estimate their efficacy as a defense, for future research.

***Physical camera discipline.*** Successful virtual theft requires that the mobile phone, in physical control of the user, be in a position to take photographs. Organizational policies against possession of camera-containing devices — such policies are commonly in place in sensitive areas — and personal discipline can both serve to prevent disclosure of sensitive images despite the capability of malicious apps.

# 5   Related Work

***Sensory malware.*** The prevalence of malware on mobile platforms is the focus of considerable research. Recent work shows more than 1,200 active variants of malware in the Android ecosystem that are largely impervious to attempts to prevent their exploitation [36]. An emerging threat is that of sensory malware, which works by leveraging the varied sensors of mobile devices. Demonstrations include using microphones to obtain dialed or spoken account information [30] and keyloggers that operate via structure-borne vibrations measured through an accelerometer [21, 22, 27]. Despite recent attention given to sensory malware, there is a notable absence of examples that recruit the camera. One example is the Stealthy Video Capturer [35]. This approach, while now dated and not reflective of the current generation of mobile devices, does present a clever method to collect image data in a stealthy manner. This type of system applied to modern Android devices would bear resemblance to the data collection segment of PlaceRaider. Where PlaceRaider differs is that we seek sampling of image data at a lower rate (0.5 frame per second vs. video frame rates) and for longer periods of time. Further, our efforts to reduce data and generate unified structures (3D models) to describe personal spaces further differentiate PlaceRaider as previous work did not attempt to make sense of the data that is collected.

***Imaging privacy attacks.*** Imaging attacks are unique in that they require a line of sight — direct or indirect via diffraction, reflection, and refraction — between the attacker and the target source (victim). In standard parlance, "standoff" refers to the physical distance between the attacker and target source. The preponderance of documented imaging attacks offer varying standoff distances, but are asymptotically limited to physical visual limitations. These examples of visual attacks on privacy require dedicated, and often advanced, imaging devices. One category of these attacks seeks to spy on active electronic displays. This can be accomplished with photomultipliers [17] or long focal length optics that can even extract information from reflections of the source image [1, 2]. Keylogging via imaging can be accomplished using techniques similar to the aforementioned attacks [29] or with thermal cameras [23]. A particularly unique visual attack can physically copy keys using images of target keys [18]. While all of these attacks are compelling, they require use of methods where attackers (or their specialized hardware) are in optical range (the physical vicinity) of the victim. PlaceRaider has a unique property in that it allows standoff distances limited only by the reach of the Internet, yet remains intimately close to the victim by leveraging the victim's own camera and sensors.

***Defenses.*** Several security enhancements have been proposed for Android. We encourage the reader to review Bugiel et al.'s report [4] for a thorough review of these approaches. In general, these approaches can check permissions at installation time [24, 11], monitor runtime permission usage [26], regulate inter-application communication to defeat confused deputy attacks [11, 9, 13, 5], mitigate covert channels [4], and track the flow of sensitive data [10, 28].

As detailed in Section 2, PlaceRaider can be deployed in a legitimate-looking camera application as a Trojan, as they share like permissions, and does not rely on covert communication. As such, there is no need for escalating privileges, so the bulk of existing defenses aimed at enforcing granted permissions offer little protection against PlaceRaider. However, context sensitive permission checks during runtime might be an effective defense against Trojans attempting to steal private data. Porscha [7] can provide such granular control for content that originates external to the phone, but does not provide protection provisions for data generated by the phone as would be necessary for PlaceRaider. Systems like CRePE [25] can provide greater levels of protection but cannot work without modifications to existing Android middle-ware.

Cai's framework for defending against sensory malware attacks offers several promising methods to thwart malicious apps [6].

Because PlaceRaider's behavior is comparable to legitimate camera apps and only requires communication to the PlaceRaider command and control server, this defensive approach would be inadequate.

The sole behavioral differences between PlaceRaider and legitimate camera applications from the system's perspective are the rate of data collection and frequent transmission of images to an off-board server. This might make PlaceRaider subject to detection via Cai's proposed information flow tracking, but ceasing the function of PlaceRaider on the basis of its behavior would also work to keep many legitimate camera apps from operating.

VirusMeter [19] is an application that seeks to identify malicious applications through their anomalous power profiles. This approach would only be effective against PlaceRaider if the rate of activity was high enough to result in an anomalous power profile.

# 6 Conclusion

In this paper we introduce a new general threat to the privacy and physical security of smartphone users that we call virtual theft. We conceptualize a mode of attack where opportunistically collected data is used to build 3D models of users' physical environments. We demonstrate that large amounts of raw data can be collected and define novel approaches that can be used to improve the quality of data that is sent to the attacker. We offer PlaceRaider, an implementation of a virtual theft attack and through human subject studies demonstrate that such attacks are feasible and powerful. An ancillary contribution is our work addressing the use of existing Structure from Motion tools in indoor spaces with poor quality images.

# Acknowledgment

This material is based upon work supported by the National Science Foundation under Grant No. 1016603. Any opinions, findings, and conclusions or recommendations expressed in this material are those of the author(s) and do not necessarily reflect the views of the National Science Foundation. We thank John McCurley for his editorial help.